 \documentclass[12pt,preprint, epsfig]{aastex}

 \received{RECEIPT DATE} \revised{REVISION DATE} \accepted{ACCEPT
 DATE} \cpright{type}{year} \journalid{VOL}{JOURNAL DATE}
 \articleid{START PAGE}{END PAGE} \paperid{MANUSCRIPT ID}
 \cpright{TYPE}{YEAR} \ccc{CODE}

 \lefthead{De Medeiros et al.}
 \righthead{The Rotation of Binary Systems with evolved Components}

\begin{document}

 \title{The Rotation of Binary Systems with Evolved Components}

 \author{J. R. De Medeiros (renan@dfte.ufrn.br), J. R. P. Da Silva
 (ronaldo@dfte.ufrn.br) and M. R. G. Maia (mrgm@dfte.ufrn.br)}
 \affil{Departamento de F\'{\i}sica, Universidade Federal do Rio
Grande do Norte, 59072--970, Natal -- RN, Brazil}


\begin{abstract}

In the present study we analyze the behavior of the rotational
velocity, $v\sin~i$, for a large sample of 134 spectroscopic
binary systems with a giant star component of luminosity class
III, along the spectral region from middle F to middle K. The
distribution of $v\sin~i$ as a function of color index (\bv) seems
to follow the same behavior as their single counterparts, with a
sudden decline around G0III. Blueward of this spectral type,
namely for binary systems with a giant F--type component, one sees
a trend for a large spread in the rotational velocities, from a
few~km~s$^{-1}$  to at least 40~km~s$^{-1}$. Along the G and K
spectral regions there is a considerable number of binary systems
with moderate to moderately high rotation rates. This reflects the
effects of synchronization between rotation and orbital motions.
These rotators have orbital periods shorter than about 250 days
and circular or nearly circular orbits. Except for these
synchronized systems, the large majority of binary systems with a
giant component of spectral type later than G0III are composed of
slow rotators.

\end{abstract}


\section{Introduction}

Rotation is one of the most important observable physical
parameters in stellar astrophysics. Such a parameter can provide
fundamental constraints for models of stellar evolution, as well
as important information on the link between surface rotation and
stellar atmospheric phenomena. The behavior of the rotational
velocity for single evolved stars of luminosity class III is now
well established (De Medeiros and Mayor, 1989, 1991; Gray, 1989).
For this luminosity class, there is a  sudden decline in the
rotational velocity around the spectral type G0III, which
corresponds to (\bv)~$\sim$~0.70. Blueward of this spectral type,
namely for F type single giants, rotational velocity scatters over
a wide range of values, from about 2 to 180 km s$^{-1}$, whereas
redward, namely for G and K--type single giants, stars are
essentially slow rotators and rotation rates greater than
5~km~s$^{-1}$ are unusual. On the basis of  the analysis of
kinematic--age relation, De Medeiros and Mayor (1991) have shown
that the root cause for this discontinuity in rotation seems to be
a mixing in ages associated with the rapid evolution of giant
stars into the Herztsprung Gap. These observational results have
been discovered from a large rotational and radial-velocity survey
of about 1100 giants (De Medeiros and Mayor 1999) accomplished
with the CORAVEL high-resolution spectrometer (Baranne et al.
1979) at the Haute Provence Observatory, in France, and at the
European Southern Observatory in Chile.

We have now examined the complete survey for the spectroscopic
binary systems containing a giant component of luminosity class
III. The main challenge in the study of rotation in binary systems
with evolved components is to establish the extent of the
synchronization between rotational and orbital motion along the
giant branch.
Tidal theory (e.g., Zahn 1977) predicts that, in
late type binary systems, viscous dissipation of time dependent
tidal effects should produce synchronization between rotation and
stellar orbital motion, as well as circularization of the orbit of
the system. The most simple way to test such effects consists of
the determination of precise rotational velocities for a large
sample of binary systems with giant components, presenting a wide
variety of values of orbital parameters. On the basis of tidal
predictions, Middelkoop and Zwaan (1981) have suggested that most
late type giants in close binary systems, with orbital periods
shorter than about 80 days, rotate in synchronization with
revolution, because most of these close binary systems have
circular orbits. Such a tendency for synchronization in late type
binary systems  was also observed by Giuricin et al. (1984).
Mermilliod and Mayor (1992) have found that, in open clusters,
binaries containing a giant have circularized orbits at orbital
periods shorter than about 250 days. More recently, Boffin et al.
(1993) have deduced a circularization cut off period of about 70
days from an eccentricity--period diagram for a large sample of
field binary systems containing late type giants.

In the present work, we study the behavior of the rotational
velocity $v\sin~i$ as a function of the color index (\bv), for a
large sample of binary systems with a giant component of
luminosity class III. We also analyze the link between $v\sin~i$
and the orbital parameters eccentricity and orbital period.

\section{The observational data}

The entire sample for the present work is composed of 134 single
lined spectroscopic binary systems, SB1, with spectral types
between F5III and K5III, mostly from the Bright Star Catalogue
(Hoffleit and Jaschek 1982; Hoffleit et al. 1983), including 73
binaries for which the orbits are known from the literature.
Observations were carried out with the CORAVEL spectrometer
(Baranne et al. 1979) mounted on the 1.0~m Swiss telescope at the
Haute Provence Observatory, France, and on the 1.44~m Danish
telescope at the European Southern Observatory, Chile, between
March 1986 and January 1994. For the determination of the
projected rotational velocity $v\sin i$, we have applied a
standard calibration (Duquennoy et al. 1991; De Medeiros and Mayor
1999) which takes into account a varying broadening mechanism, as
a function of color, probably related to turbulent motions and/or
magnetic fields. Nevertheless, let us recall that such calibration
is an extension of the method developed by Benz and Mayor (1981,
1984) which is based on the cross correlation technique (Griffin
1967). For the observational procedure, calibration process and
error analysis, the reader is referred to De Medeiros and Mayor
(1999). A comparison of our measurements, obtained by the cross
correlation technique, with those acquired with the
Fourier--transform technique by Gray (1989) for a common sample of
84 single and binary giant and subgiant stars, gives excellent
agreement, with a typical rms of the rotational velocity
differences of about 1.1~km~s$^{-1}$, indicating a precision of
1.0~km~s$^{-1}$ for our $v\sin~i$ measurements (De Medeiros and
Mayor 1999). This excellent agreement between the $v\sin~i$ values
from CORAVEL and Fourier transform is confirmed for the binary
stars, when they are analyzed separately. By using the $v\sin~i$
data given in Table 1 for 15 stars of the present sample, we have
found a rms of the rotational velocity differences of
0.61~km~s$^{-1}$.

\placetable{tab1}

The entire sample for the present work, with rotational velocity
and orbital parameters, when the latter are available, is listed
in Table 3. For those stars with no available orbital parameters
Table 3 gives also for each star, the number $N$ of observations;
the radial velocity dispersion (rms) $\sigma$; the uncertainty
$\epsilon$ on the mean radial velocity, given by
max~$(\epsilon_1/\sqrt{N}, \sigma/\sqrt{N})$, where $\epsilon1$ is
the typical error for one single radial velocity measurement; the
time span $\Delta{T}$ of the observations and the
$\sigma/\sigma_n$, a factor indicating how many times their radial
velocity dispersion (rms) exceeds the radial velocity noise
expected for single stars.

For the stars with no orbital parameters listed in Table 3, one
important question concerns the nature of their radial velocity
variation. Is such a variability reflecting the presence of a
dynamical companion, or is intrinsic to the star, reflecting
either rotational modulation by surface features or nonradial
pulsations? In this context, different studies (e.g. Hatzes and
Cochran 1993; Frink et al. 2001) have revealed intrinsic radial
velocity variability in K giant stars with amplitudes ranging from
0.2 to 0.4 ~km~s$^{-1}$. Using the Maximum Likelihood approach we
have analyzed the behavior of the factor $\sigma/\sigma_n$ defined
above, to estimate if the radial velocity variability of such
stars is likely indicating a dynamical companion. For this, we
have estimated the expected radial velocity noise $\sigma$ for
single giants, from intrinsic contributions with typical amplitude
of 0.4~km~s$^{-1}$. This amplitude value was added in quadrature
with the 0.3~km~s$^{-1}$, the typical precision of a single radial
velocity measurement by CORAVEL (see Duquennoy et al. 1991),
giving a radial velocity noise $\sigma_n$ of 0.50~km~s$^{-1}$.
With this expected value of $\sigma_n$ in hand, we have obtained
the factor $\sigma/\sigma_n$ listed in Table 3. Afterwards, we
have estimated the $\sigma/\sigma_n$ for a sample of 641 G and K
single giant stars from De Medeiros and Mayor (1999). The
distributions of the factor $\sigma/\sigma_n$ for single giants
and for the stars with no orbital parameters in Table 3 are
displayed in Fig. 1, both very well fitted by a gaussian function.
For a best presentation of the distributions, the
$\sigma/\sigma_n$ values in Fig. 1 were normalized by the highest
$\sigma/\sigma_n$ for each group of stars, namely 1.9 and 35.84
for single giants and for the stars with no orbital parameters,
respectively. In addition, the frequency of single stars was
reduced by a factor 5. From the Maximum Likelihood statistics we
obtained a $<\sigma/\sigma_n>$ of $0.33\pm0.36$ and $6.32\pm4.67$
for single giants and for the stars with no orbital parameters in
Table 3, respectively. From this analysis we estimate that the
threshold to indicate a dynamical companion is around a
$\sigma/\sigma_n$ of 0.69. Stars with no orbital parameters listed
in Table 3 present a $\sigma/\sigma_n$ higher that such threshold
value, indicating that their radial velocity variabilities reflect
very probably a binarity behavior.

%
\begin{figure}
\figurenum{1} \epsscale{1.0} \plotone{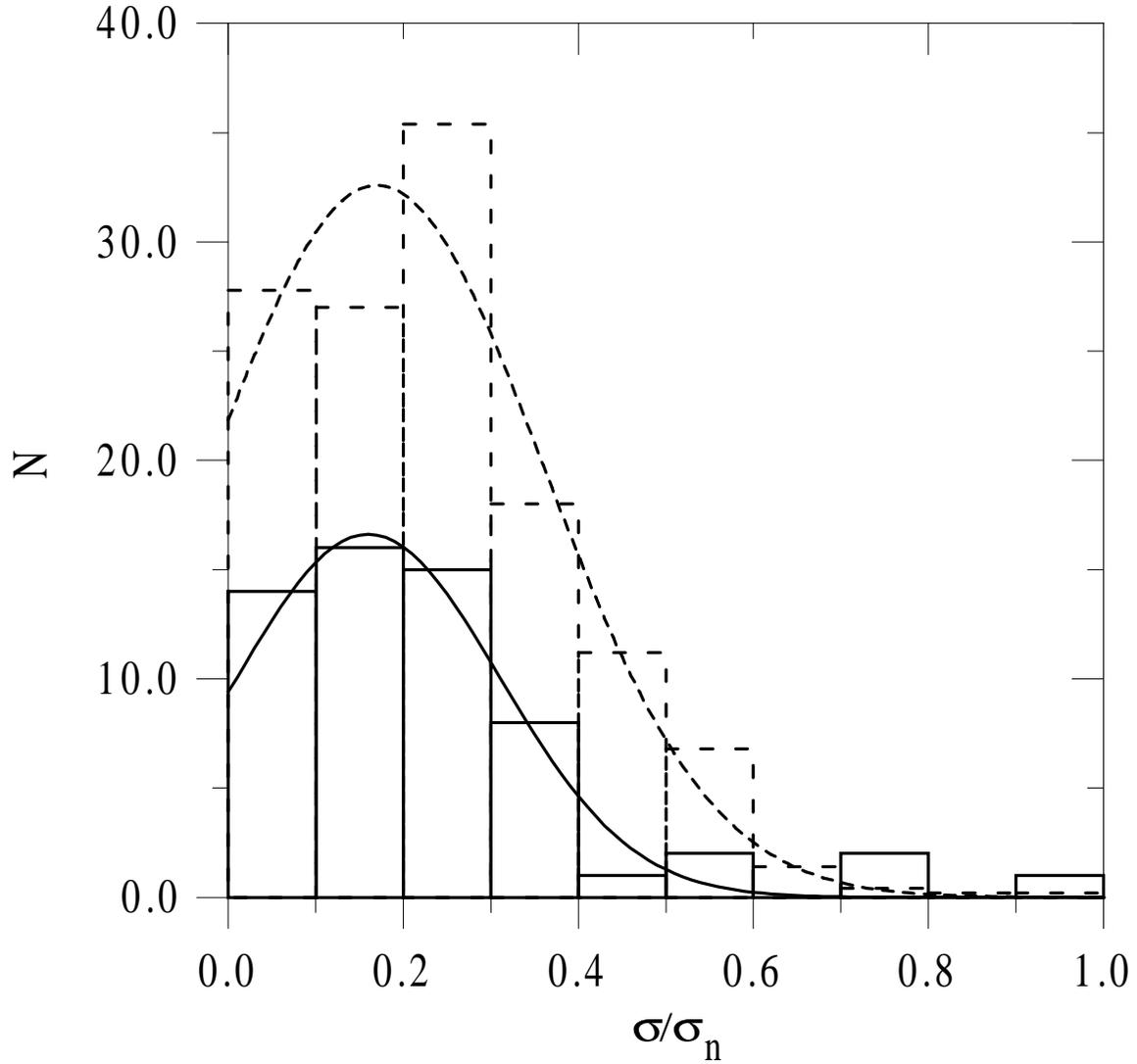}
\figcaption{The distribution of the factor $\sigma/\sigma_n$ for single giants stars
(dashed line) and for the binary systems with no orbital parameters (solid line). For
a best presentation of the distributions, $\sigma/\sigma_n$ values and the number of
single giants were normalized as explained in the text.
\label{fig1}}
 \end{figure}

It is important to underline that in this paper we are analyzing
only the binary
systems presenting a SB1 behavior on the basis of our CORAVEL
observations. Double lined systems, SB2, and binaries presenting a
composite spectrum will be discussed in a forthcoming work. We
have excluded from the sample all those systems classified in the
literature with a composite spectrum, in spite of the fact that,
for some systems, the CORAVEL observations have shown a SB1
behavior.

\section{Results and Discussion}

The main results of the present work are displayed in
Figures~2a,b and \ref{fig3}, where we show the distribution of the
rotational velocities as a function of the color index (\bv) and
orbital parameters. Several important features are well marked.

%
\begin{figure}
 \figurenum{2} \epsscale{1.0} \plottwo{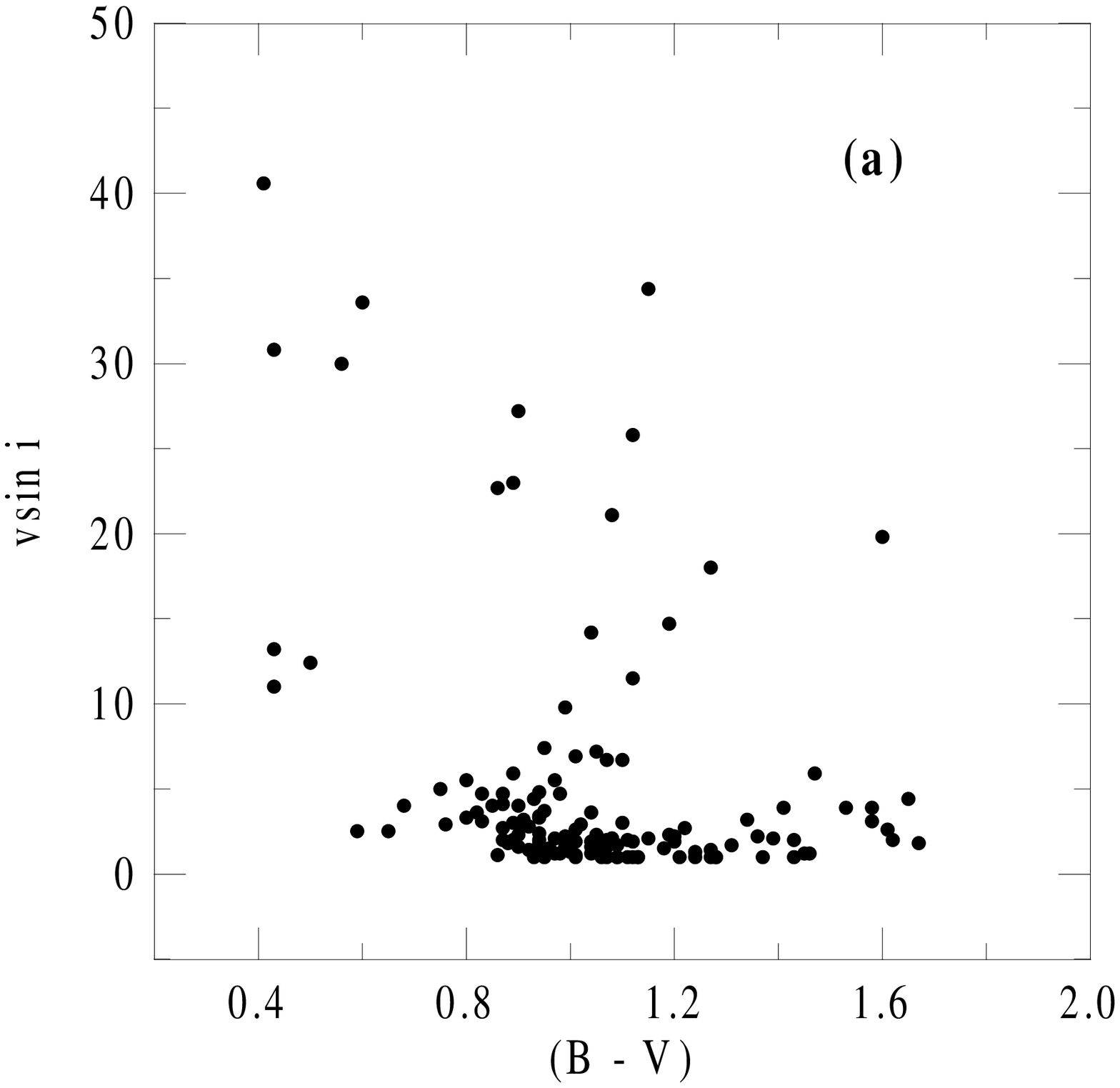}{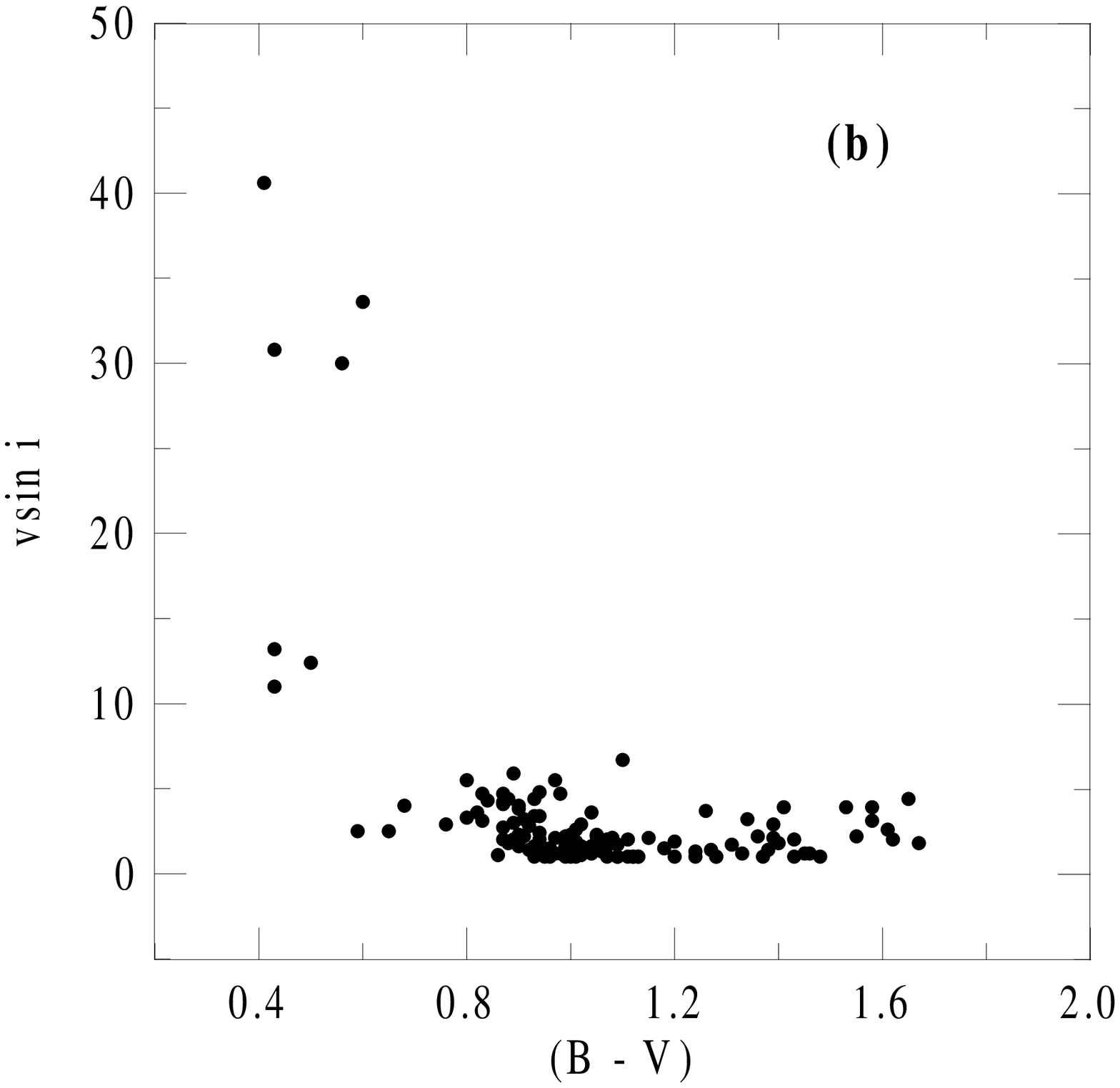}
\figcaption{The distribution of the rotational velocity $v\sin~i$
for binary systems with evolved component. (a) For all the stars
listed in Table 3. (b) For systems with G-- or K--type component
but with orbital period longer than 250 days and all the systems
with F--component. \label{fig2}}
 \end{figure}

Figure \ref{fig2}a shows the general trend of the distribution of
the rotational velocity $v\sin~i$ as a function of the color index
(\bv) for spectroscopic binary systems with evolved component,
with a tendency for a sudden decline around (\bv)~$\sim$~0.70,
corresponding to the spectral type G0III. This behavior seems to
follow that of the rotational velocity presented by their single
counterparts. Blueward of the spectral type G0III there is a
spread in the values of rotational velocity, ranging from a few km
s$^{-1}$ to at least 40~km~s$^{-1}$, whereas redward of this
spectral type, namely along the G and K spectral regions, the
great majority of binary systems rotate very slowly, following the
same trend observed for single giants. Let us recall that the mean
rotational velocity $v\sin~i$ for single G and K type giants
ranges typically from about 6~km~s$^{-1}$ at G1III, to about
3~km~s$^{-1}$ at G5III and to about 2~km~s$^{-1}$ along the
spectral region from G8III to K7III (De Medeiros et al. 1996).
Nevertheless, twelve out of 124 binary systems located redward of
G0III showing moderate to moderately high rotation rates,
typically stars with $v\sin~i$ higher than 6~km~s$^{-1}$. For a
more consistent analysis on the extent of the decline in $v\sin~i$
shown in Figure \ref{fig2}a, one should inquire about the root
cause of the enhanced rotation presented by such binary systems.
By analyzing the orbital parameters given in Table 3, we find that
the G and K binary systems with enhanced rotation have an orbital
period shorter than about 250 days and a circular or nearly
circular orbit, such a value pointing for the critical period of
synchronization between axial rotational motion and orbital
revolution due to tidal effects. The critical period of
synchronization around 250 days found on the basis of a large
sample of field binary systems, sounds an interesting result
because it follows that found by Mermilliod and Mayor (1992) from
a sample of 88 binary systems with an evolved component in open
clusters. In this context, one can conclude that the enhanced
rotation of binary systems along the G and K spectral regions
results from the synchronization between rotation and orbital
motions. Hence, the moderate to moderately high rotation shown by
these synchronized binary systems reflects angular momentum that
has been drawn from the orbital motion. The orbit acts as a source
of angular momentum to replace that lost via stellar wind, during
the evolution along the giant branch. It is clear that for some
systems the synchronization may increase their observed rotational
velocity by more than about 15 times the mean rotational velocity
observed for single giant stars at a given spectral type.
Following Zahn (1966, 1977), in binary stars possessing a
convective envelope, tidal effects become important if the
convective region occupies a substantial fraction of the star, and
if convection transports most of the energy flux. According the
tidal theory, in the components  of a late type binary system, the
tidal wave, which is due to the gravitational interaction, is
slightly lagging because a small fraction of its kinetic energy is
converted into heat, as required by the second principle of
thermodynamics. The system evolves towards its state of minimum
kinetic energy, in which the orbit is circular, the spin axes are
aligned and perpendicular to the orbital plane and the rotation of
both stars of the system is synchronized with the orbital motion.
Such synchronism is predicted to be completed before orbits become
circular (Zahn 1977), unless the spin angular momentum is
comparable to the orbital angular momentum (Zahn 1977, Hut 1980).
Since evolutionary models predict that the rapid increase of the
thickness of the convective envelope coincides with the late F
spectral region (e.g., Maeder and Meynet 1994), our finding
presents strong evidence that the synchronization is achieved once
stars arrive in the late F to early G spectral regions.

At this point one should ask about the nature and the extent of
the decline in rotation around G0III, suggested by Figure
\ref{fig2}a. Is it really paralleling the rotational discontinuity
at G0III observed in single giant stars? In fact we see two
possible regimes. First, the synchronization is achieved only for
stars along the G to K spectral regions (let us recall that the
extent of the convective envelope and the effectiveness of
convection in transporting energy flux become important only when
the star evolves up the giant branch). In this context we can
consider that binary systems in the F spectral region, in
particular those with early to middle F--type component, are not
synchronized, their rotation reflecting only the normal evolution
of the star with no effects from binarity. Following this
scenario, in Figure \ref{fig2}b we represent the rotational
velocity $v\sin~i$ as a function of the color index (\bv) for
binary systems with G-- or K--type component, but for orbital
periods longer than 250 days and all systems with an F--type
component. We have also excluded from this figure the G and K
binary systems with significant enhanced rotation in respect to
the mean, HR~2054, HR~2145 and HR~2376, for which no orbital
parameters are available in the literature, but their large radial
velocity range, respectively 9.58~km~s$^{-1}$, 10.59~km~s$^{-1}$,
and 19.19~km~s$^{-1}$, indicates for short or moderately short
orbital periods. The sudden decline in $v\sin~i$ around (\bv) =
0.70 is now quite clear. Of course, low rotator binary systems
situated along the G and K spectral regions should be mostly
nonsynchronized systems. Consequently, they should have orbital
periods longer than the cut--off period of about 250 days and
rather non circular orbits. Nevertheless, a few systems in Table
3, like HR~373 and HR~503, have nearly circular orbits with
periods shorter than 250 days and low rotation. Perhaps their true
rotation rates are high but $v\sin~i$ is suppressed by low
inclination of their rotation axis. Second, the F-- type stars
with short orbital period like HR~765, and HR~2264 would have also
reached a stage where convective envelope is sufficiently
developed for synchronization, in spite of their middle--F
spectral types. Following such a scenario we can also ask about
the nature of the additional F-type stars represented in Figure
\ref{fig2}b but with no orbital parameters available in the
literature. If these stars are also in synchronization there is no
supporting evidence for a rotational discontinuity in binary
systems following that observed for single giants, as proposed in
the first scenario above discussed. To decide between these two
scenarios it is necessary, first, to determine orbital parameters
for all the F--type stars composing the present sample and,
second, to determine the rotational velocity for a larger number
of F--type stars.

In Figure \ref{fig3} we show the behavior of the rotational
velocity as a function of the orbital period. Stars are separated
in two groups: those with circular or nearly circular orbits,
namely an eccentricity $e$ lower than about 0.10, and those with
non circular orbits, namely an eccentricity $e$ higher than 0.10.
This figure shows clearly that the binary systems with enhanced
rotation, typically $v\sin~i$ greater than about 6~km~s$^{-1}$,
have an orbital period shorter than about 250 days and circular or
nearly circular orbit. Two additional trends are clearly observed
in this figure: first, high rotators with orbital period longer
than 250 days seems to be unusual and second, there is an absence
of slow rotators with very short orbital period. These results are
consistent with the expected features for the synchronization
process in binary systems with evolved components. While the great
majority of binary systems with enhanced rotation present a
circular or nearly circular orbit, one star represented in Figure
\ref{fig3}, HR~407, appears to violate the general rule. This
star, in spite of the very short orbital period of 5.4291 days,
has a F5III spectral type which may indicate that, at this
evolutionary stage, the extent of its convective envelope and the
effectiveness of convection in transporting energy flux are not
yet well developed, indicating that tidal effects are not yet
enough to produce a circular or nearly circular orbit. On the
basis of this reasoning the deviating behavior of HR~407 in the
$v\sin i$--period--eccentricity plane is easily understood.
Naturally, enhanced rotation is not a general property among
binary systems with circularized orbits. As shown in Figure
\ref{fig3}, binary systems with circular or nearly circular orbits
present a wide range of $v\sin~i$  values, from a few~km~s$^{-1}$
to about 40~km~s$^{-1}$.

\begin{figure}
 \figurenum{3} \epsscale{1.0} \plotone{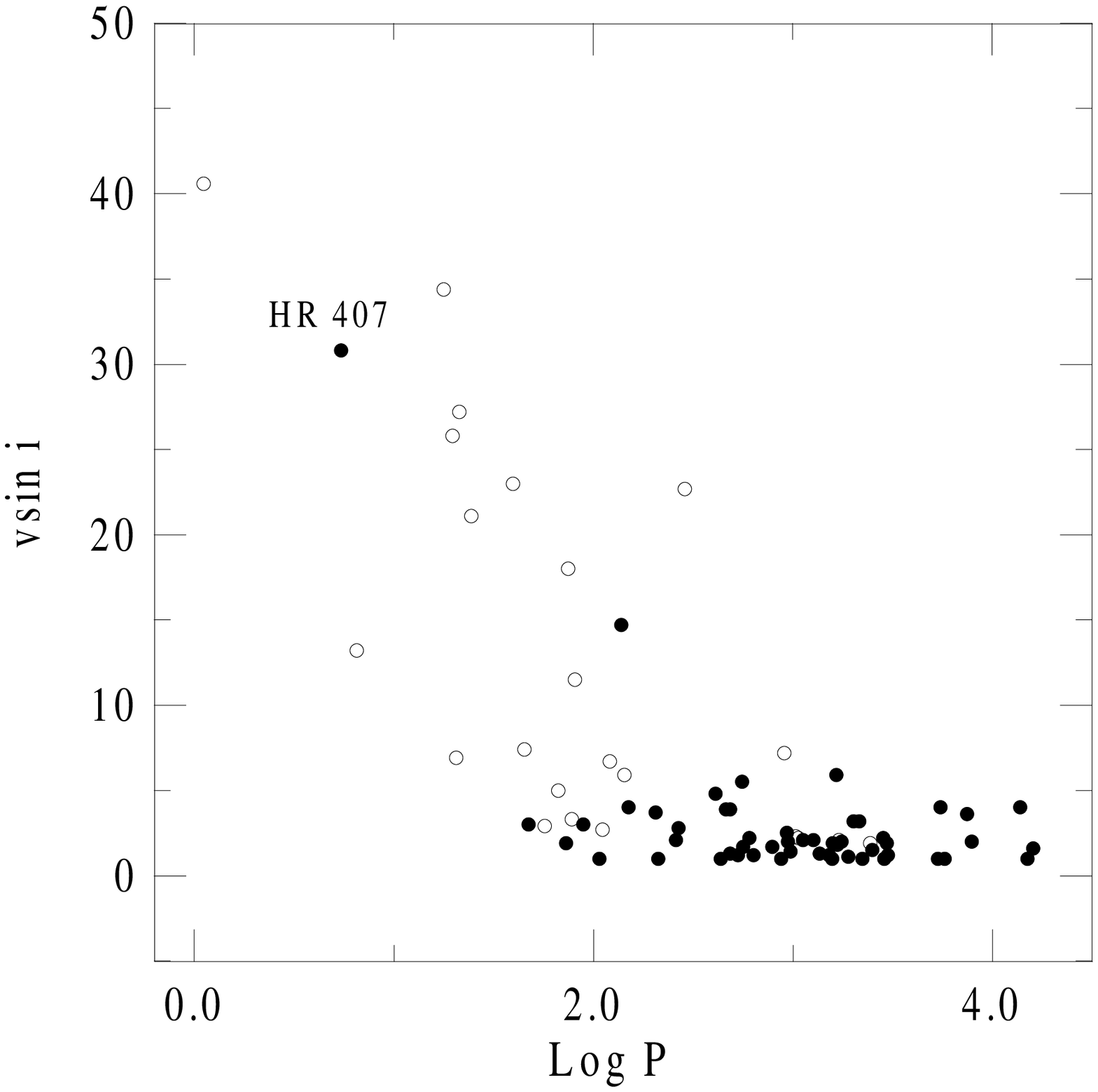}
\caption{The distribution of the rotational velocity $v\sin~i$ as
a function of the orbital period, for the binary systems listed in
Table 3 with available orbital parameters. Binary systems with a
circular or nearly circular orbit, ($e~\leq~0.10$) are represented
by open symbols whereas the systems with eccentric orbits,
($e~>~0.10$) are represented by solid symbols. The deviating
behavior of the star HR~407 is discussed in the text\label{fig3}}
\end{figure}

We have also analyzed the behavior of the rotational velocity,
$v\sin~i$, versus the mass function, $f(m)$, searching for
particular features between rotation and stellar mass. Clearly,
the mass function per se is not a very useful parameter because of
its rather complex dependence on the inclination angle $i$ of the
orbital plane with respect to the plane of the sky. The relevance
of such analysis rests on the fact that the mass function, $f(m)$,
has a direct dependence on the orbital period, $P$, and
eccentricity, $e$, namely on $P$ and $(1 - e^{2})$, respectively.
Nevertheless, no significant feature resulted from this analysis.

\section{Conclusions}

Precise rotational velocities, $v\sin~i$, are given for a large
sample of 134 single lined binary systems with an evolved
component of luminosity class III. For these binary systems the
distribution of rotation rates as a function of color index (\bv)
presents a behavior that seems to parallel the one found for their
single counterparts. Namely, there is a sudden decline in rotation
around the spectral type G0III, as is the case for single giants.
Binary systems located blueward of the spectral type G0III,
typically those systems with a color index (\bv) lower than about
0.70, present a large spread in the values of rotational velocity,
from a few~km~s$^{-1}$ to at least 40~km~s$^{-1}$. In addition, we
have found that along the G and K spectral regions there is a
considerable number of moderate to moderately high rotators
reflecting, clearly, the effects of synchronization between
rotation and orbital motions. These stars have orbital periods
shorter than about 250 days as well as circularized or nearly
circularized orbits. For a given spectral type, the process of
synchronization increases the observed rotational velocity of the
binary systems up to about 15 times the mean rotational velocity
of single giants. Except for these synchronized systems, the
majority of binary stars later than spectral type G0III is
essentially composed of slow rotators, following the same behavior
as single giants. A few binary systems present synchronization
characteristics and low rotation but probably their $v\sin~i$ is
suppressed by low inclination of their rotation axis. We have also
observed that enhanced rotation is not a general property among
binary systems with circularized orbit. Admittedly, in the present
study the number of binary systems with F--type component is very
scarce and one should be cautious with the proposed rotational
discontinuity around the spectral type G0III. In addition the
extent of tidal effects along the F spectral region is not yet
established. In this context only the determination of rotational
velocity and orbital parameters for a larger sample of binary
systems with F--type component could confirm such a discontinuity
on more solid basis. Finally, we would like to point out that for
the large majority of stars discussed in this work the duplicity
is established on very solid grounds. However for 58 stars
additional measurements of radial velocity are undoubtedly
necessary to establish their orbital parameters. Nevertheless, it
is important to underline that the discussion of the rotational
velocity for binary systems with evolved component carried out in
the present work is not hampered by this fact because, in
particular, the large majority of stars with no available orbital
parameters are slow rotators.

\acknowledgments

We acknowledge the Geneva Observatory for the important amount of
observing time awarded for this work. It is a pleasure for J. R.
M. to thank warmly Prof. Michel Mayor for his generous help all
along this research and all colleagues at the Geneva Observatory
who made the time in Geneva pleasant and productive. We express
our thanks to the Referee for his careful reading and constructive
comments on the original version of this paper. We are grateful to
the CNPq and CAPES Brazilian Councils for continued financial
support.


\clearpage

\begin{deluxetable}{rrrr}
 \footnotesize
 \tablecaption{Comparison between $v\sin~i$, in~km~s$^{-1}$, obtained by CORAVEL (COR) and
 Fourier transform technique (Gray).\label{tab1}}
 \tablenum{2}
 \tablewidth{0pt}
 \tablehead{ \colhead{HR} & \colhead{HD} &
 \colhead{$v\sin~i$}(COR) & \colhead{$v\sin~i$}(Gray)\\}

\startdata


 373  &   7672    &       2.9  & 4.5 \\
 1030 &   21120   &       5.9  &          4.8 \\
 1373 & 27697   & 1.2  &          2.5 \\
 3112 &   65448   & 2.5  &          2.4 \\
 4301 &   95689   & 1.6  &          2.6 \\
 5161 & 119458  &       4.0  &    4.9 \\
 5681 &   135722  & 1.2  &          1.1 \\
 6148 &   148856  & 4.8  & 3.4 \\
 6239 &   151627  &       4.1  &          4.7 \\
 6322 & 153751  &      23.0  &          24.0 \\
 7995 &   198809  &       4.7  & 5.9 \\
 8359 &   208010  &       3.3  &          4.0 \\
 8442 & 210220  & 1.8  &          3.4 \\
 8819 &   218658  & 5.5  &          4.7 \\
 8923 &   221115  &       1.5  & 3.1 \\

\enddata
\end{deluxetable}


\tablenum{3}
 \clearpage
 \tablewidth{0pt}
 \begin{deluxetable}{rrrcrrrrrrrrrr}
 \tabletypesize
 \scriptsize
 \tablecolumns{13}
 \tablecaption{Stellar parameters for binary systems with evolved
 components.\label{tab2}}

 \tablehead{HR & HD & $\bv$    & Spectral & $v\sin~i$ & $P$   &
 $e$ & N & $\sigma$ & $\epsilon$ & $\Delta{T}$ &
 $\sigma/\sigma_n$ &  ref. \\
                      &
                      &                        &
type        &  kms$^{-1}$  & days &  &  & kms$^{-1}$
                      &   kms$^{-1}$    &  days
                      &       & }
\startdata
 3    & 28     & 1.04 & K1III         &  1.9 &   72.93 &    0.27 & \nodata & \nodata & \nodata & \nodata & \nodata &       2 \\
 40   & 895    & 0.65 & G0III         &  2.5 &  932.22 &   0.832 & \nodata & \nodata & \nodata & \nodata & \nodata &      44 \\
 165  & 3627   & 1.28 & K3III         &  1.0 &   15000 &    0.34 & \nodata & \nodata & \nodata & \nodata & \nodata &       2 \\
 216  & 4526   & 0.94 & G8III         &  1.7 & \nodata & \nodata &       2 &    9.96 &    7.04 &     309 & 19.92    & \nodata \\
 360  & 7318   & 1.04 & K0III         &  3.6 &    7473 &   0.816 & \nodata & \nodata & \nodata & \nodata & \nodata &      27 \\
 373  & 7672   & 0.90 & G5IIIe        &  2.9 & 56.8147 &    0.04 & \nodata & \nodata & \nodata & \nodata & \nodata &       2 \\
 407  & 8634   & 0.43 & F5III         & 30.8 &  5.4291 &    0.38 & \nodata & \nodata & \nodata & \nodata & \nodata &       2 \\
 469  & 10072  & 0.89 & G8III         &  2.0 &    7581 &   0.368 & \nodata & \nodata & \nodata & \nodata & \nodata &      29 \\
 503  & 10588  & 0.94 & G8III--IV     &  3.3 & 78.0073 &    0.02 & \nodata & \nodata & \nodata & \nodata & \nodata &       2 \\
 549  & 11559  & 0.94 & K0III         &  1.8 &  1672.4 &    0.18 & \nodata & \nodata & \nodata & \nodata & \nodata &      36 \\
 645  & 13530  & 0.93 & G8III:v       &  1.0 & 1575.48 &  0.8815 & \nodata & \nodata & \nodata & \nodata & \nodata &       7 \\
 731  & 15596  & 0.90 & G5III--IV     &  1.6 & \nodata & \nodata &       2 &    1.20 & 0.85    & 301     & 2.40    & \nodata \\
 738  & 15755  & 1.07 & K0III         &  1.0 & \nodata & \nodata &       2 &   13.17 & 9.31    & 285     & 26.34    & \nodata \\
 754  & 16161  & 0.87 & G8III         &  2.7 & \nodata & \nodata &       2 &    6.73 & 4.76    & 302     & 13.46    & \nodata \\
 765  & 16246  & 0.41 & F6III         & 40.6 &1.109526 &   0.062 & \nodata & \nodata & \nodata & \nodata & \nodata &      42 \\
 831  & 17484  & 0.43 & F6III--IV     & 11.0 & \nodata & \nodata &       2 &    2.04 &    1.44 &     302 &  4.08   & \nodata \\
 1023 & 21018  & 0.86 & G5III         & 22.7 & 287.201 &     0.0 & \nodata & \nodata & \nodata & \nodata & \nodata &      39 \\
 1030 & 21120  & 0.89 & G6IIIFe--1    &  5.9 &  1654.9 &   0.263 & \nodata & \nodata & \nodata & \nodata & \nodata &      38 \\
 1198 & 24240  & 1.05 & K0III         &  2.3 & \nodata & \nodata &       2 &    3.75 &    2.65 &     419 & 7.50 & \nodata \\
 1304 & 26659  & 0.87 & G8III         &  4.7 & \nodata & \nodata &       5 &    1.58 &    0.71 &    2037 & 3.16 & \nodata \\
 1313 & 26755  & 1.09 & K1III         &  1.0 & \nodata & \nodata &       2 &    6.72 &    4.75 &     267 & 13.44 & \nodata \\
 1337 & 27278  & 0.94 & K0III         &  2.0 & \nodata & \nodata &       2 &    2.62 &    1.85 &     267 & 5.24 & \nodata \\
 1360 & 27497  & 0.92 & G8III--IV     &  1.4 & \nodata & \nodata &       4 &    3.12 &    1.56 &    3637 & 6.24 & \nodata \\
 1373 & 27697  & 0.98 & K0IIICN0.5    &  1.2 &   529.8 &    0.42 & \nodata & \nodata & \nodata & \nodata & \nodata &      35 \\
      & 28591  & 0.90 & K1III         & 27.2 & 21.2886 &   0.010 & \nodata & \nodata & \nodata & \nodata & \nodata &       7 \\
 1467 & 29317  & 1.07 & K0III         &  6.7 &     121 &    0.02 & \nodata & \nodata & \nodata & \nodata & \nodata &       2 \\
 1514 & 30138  & 0.93 & G9III         &  4.4 & \nodata & \nodata &       2 &    1.76 &    1.24 &     329 & 3.52 & \nodata \\
 1517 & 30197  & 1.21 & K4III         &  1.0 & 107.503 &   0.210 & \nodata & \nodata & \nodata & \nodata & \nodata &      32 \\
 1623 & 32357  & 1.12 & K0III         & 11.5 &   80.90 &    0.05 & \nodata & \nodata & \nodata & \nodata & \nodata &      37 \\
 1698 & 33856  & 1.19 & K3III         &  2.3 & 1031.40 &   0.098 & \nodata & \nodata & \nodata & \nodata & \nodata &       4 \\
 1726 & 34334  & 1.27 & K3III         &  1.0 &   434.8 &     0.1 & \nodata & \nodata & \nodata & \nodata & \nodata &       6 \\
 1908 & 37171  & 1.58 & K4III         &  3.9 & \nodata & \nodata &       2 &    5.78 &    4.08 &     367 & 11.56  & \nodata \\
 1970 & 38099  & 1.47 & K4III         &  5.9 &  143.03 &    0.06 & \nodata & \nodata & \nodata & \nodata & \nodata &       1 \\
 2054 & 39743  & 0.99 & G8III         &  9.8 & \nodata & \nodata &       3 &    4.87 &    2.81 &    1067 & 9.74 & \nodata \\
 2077 & 40035  & 1.00 & K0III         &  1.7 & \nodata & \nodata &       2 &    1.97 &    1.39 &     288 & 3.94 & \nodata \\
 2145 & 41380  & 1.04 & G4III         & 14.2 & \nodata & \nodata &       3 &    6.25 &    3.61 &    1079 & 12.50 & \nodata \\
 2264 & 43905  & 0.43 & F5III         & 13.2 &  6.5013 &    0.02 & \nodata & \nodata & \nodata & \nodata & \nodata &       2 \\
 2376 & 46101  & 1.60 & K0III:        & 19.8 & \nodata & \nodata &       2 &   13.48 &    9.53 &     282 & 26.96 & \nodata \\
 2506 & 49293  & 1.11 & K0IIIBa0.1    &  2.0 &  1760.9 &    0.40 & \nodata & \nodata & \nodata & \nodata & \nodata &      18 \\
 2553 & 50310  & 1.20 & K0III         &  2.2 &  1066.0 &    0.09 & \nodata & \nodata & \nodata & \nodata & \nodata &       2 \\
 2804 & 57646  & 1.61 & K5III         &  2.6 & \nodata & \nodata &       2 &    1.20 &    0.85 &     219 &  2.40 & \nodata \\
 2973 & 62044  & 1.12 & K1III         & 25.8 &  19.605 &       0 & \nodata & \nodata & \nodata & \nodata & \nodata &      41 \\
 3003 & 62721  & 1.45 & K5III         &  1.2 &  1519.7 &   0.325 & \nodata & \nodata & \nodata & \nodata & \nodata &      13 \\
 3043 & 63660  & 0.76 & G0III         &  2.9 & \nodata & \nodata &       3 &    5.15 &    2.97 &    1401 & 10.03 & \nodata \\
 3112 & 65448  & 0.59 & G1III         &  2.5 & \nodata & \nodata &       2 &    7.67 &    5.43 &     171 & 15.34 & \nodata \\
 3149 & 66216  & 1.12 & K2III         &  1.9 &  2437.8 &   0.060 & \nodata & \nodata & \nodata & \nodata & \nodata &      13 \\
 3245 & 69148  & 0.89 & G8III         &  3.0 &89.06533 &    0.19 & \nodata & \nodata & \nodata & \nodata & \nodata &       5 \\
 3360 & 72184  & 1.11 & K2III         &  1.0 & \nodata & \nodata &       2 &    3.62 &    2.56 &     410 & 7.24 & \nodata \\
 3385 & 72688  & 0.95 & K0III         &  7.4 &   45.13 &       0 & \nodata & \nodata & \nodata & \nodata & \nodata &      48 \\
 3482 & 74874  & 0.68 & G5III         &  4.0 &    5492 &    0.61 & \nodata & \nodata & \nodata & \nodata & \nodata &       2 \\
 3512 & 75605  & 0.87 & G5III         &  2.0 & \nodata & \nodata &       3 &    1.89 &    1.09 &    1440 & 3.78 & \nodata \\
 3531 & 75958  & 0.86 & G6III         &  1.1 &  1898.7 &   0.706 & \nodata & \nodata & \nodata & \nodata & \nodata &      30 \\
 3567 & 76629  & 0.98 & G8III         &  4.7 & \nodata & \nodata &       3 &    4.96 &    2.86 &    2290 & 9.92 & \nodata \\
 3627 & 78515  & 0.97 & K0III         &  2.1 & 1700.76 &   0.060 & \nodata & \nodata & \nodata & \nodata & \nodata &      38 \\
 3722 & 80953  & 1.46 & K2III         &  1.2 & \nodata & \nodata &       2 &    4.43 &    3.13 &     265 & 8.86 & \nodata \\
 3725 & 81025  & 0.75 & G2III         &  5.0 &  66.717 &     0.0 & \nodata & \nodata & \nodata & \nodata & \nodata &       2 \\
 3827 & 83240  & 1.05 & K1IIIv        &  2.2 &    2834 &   0.322 & \nodata & \nodata & \nodata & \nodata & \nodata &      20 \\
 3907 & 85505  & 0.94 & G9III         &  3.4 & \nodata & \nodata &       4 &    3.01 &    1.50 & 2309    & 6.02   & \nodata \\
 3994 & 88284  & 1.01 & K0III         &  1.9 &  1585.8 &    0.14 & \nodata & \nodata & \nodata & \nodata & \nodata &       2 \\
 4100 & 90537  & 0.90 & G9IIIab       &  4.0 &   13833 &    0.66 & \nodata & \nodata & \nodata & \nodata & \nodata &       2 \\
 4235 & 93859  & 1.12 & K2III         &  1.0 & \nodata & \nodata &       2 &    1.09 &    0.77 &     419 & 2.18 & \nodata \\
 4301 & 95689  & 1.07 & K0IIIa        &  1.6 &   16060 &    0.35 & \nodata & \nodata & \nodata & \nodata & \nodata &       2 \\
 4365 & 97907  & 1.20 & K3III         &  1.9 &    2963 &   0.420 & \nodata & \nodata & \nodata & \nodata & \nodata &      23 \\
 4427 & 99913  & 0.94 & K0III         &  2.4 & \nodata & \nodata &       2 &    3.13 &    2.21 &     320 & 6.26 & \nodata \\
 4430 & 99967  & 1.27 & K2IIICN--1    & 18.0 &  74.861 &    0.03 & \nodata & \nodata & \nodata & \nodata & \nodata &       2 \\
 4451 & 100418 & 0.60 & F9III         & 33.6 & \nodata & \nodata &       5 &    3.72 &    1.67 &    1803 & 7.44 & \nodata \\
 4593 & 104438 & 1.01 & K0III         &  1.1 & \nodata & \nodata &       2 &    1.77 &    1.25 &     243 & 3.54 & \nodata \\
 4640 & 105981 & 1.41 & K4III         &  3.9 &     461 &    0.17 & \nodata & \nodata & \nodata & \nodata & \nodata &       2 \\
 4693 & 107325 & 1.09 & K2III--IV     &  1.0 &    5792 &    0.55 & \nodata & \nodata & \nodata & \nodata & \nodata &      24 \\
 4793 & 109519 & 1.22 & K1III         &  2.7 & 110.829 &       0 & \nodata & \nodata & \nodata & \nodata & \nodata &      34 \\
 4795 & 109551 & 1.31 & K2III         &  1.7 &   561.7 &   0.262 & \nodata & \nodata & \nodata & \nodata & \nodata &      33 \\
 4815 & 110024 & 0.96 & G9III         &  1.4 &   972.4 &   0.590 & \nodata & \nodata & \nodata & \nodata & \nodata &      11 \\
 4927 & 113049 & 0.99 & K0III         &  1.6 & \nodata & \nodata &       4 &    2.88 &    1.44 &    1838 & 5.76 & \nodata \\
 5053 & 116594 & 1.06 & K0III         &  1.3 &  1366.8 &   0.193 & \nodata & \nodata & \nodata & \nodata & \nodata &      21 \\
 5161 & 119458 & 0.85 & G5III         &  4.0 &  149.72 &    0.17 & \nodata & \nodata & \nodata & \nodata & \nodata &       3 \\
 5201 & 120539 & 1.43 & K4III         &  2.0 &     944 &    0.41 & \nodata & \nodata & \nodata & \nodata & \nodata &       2 \\
 5203 & 120565 & 1.01 & G9III         &  2.6 & \nodata & \nodata &       2 &    1.24 &    0.88 &     341 & 2.48 & \nodata \\
 5321 & 124547 & 1.36 & K3III         &  2.2 &   605.8 &   0.137 & \nodata & \nodata & \nodata & \nodata & \nodata &      45 \\
 5361 & 125351 & 1.06 & K0III         &  1.0 & 212.085 &   0.574 & \nodata & \nodata & \nodata & \nodata & \nodata &      47 \\
 5520 & 130458 & 0.82 & G5III         &  3.6 & \nodata & \nodata &       3 &    2.63 &    1.52 &    2555 & 5.26 & \nodata \\
 5681 & 135722 & 0.95 & G8IIICN--1    &  1.2 & \nodata & \nodata &      11 &    2.98 &    0.90 &    5213 & 5.96 & \nodata \\
 5692 & 136138 & 0.97 & G8IIIaBa0.3   &  5.5 & \nodata & \nodata &       4 &    5.28 &    2.64 &     996 & 10.56 & \nodata \\
 5769 & 138525 & 0.50 & F6III         & 12.4 & \nodata & \nodata &       2 &    6.64 &    4.70 &     215 & 13.28 & \nodata \\
 5802 & 139195 & 0.95 & K0III:CN1B    &  1.0 &    5324 &   0.345 & \nodata & \nodata & \nodata & \nodata & \nodata &      25 \\
 5826 & 139669 & 1.58 & K5III         &  3.1 & \nodata & \nodata &       2 &    1.28 &    0.90 &     708 & 2.56 & \nodata \\
 5835 & 139906 & 0.83 & G8III         &  3.1 & \nodata & \nodata &       5 &    5.43 &    2.43 &    2514 & 10.86 & \nodata \\
 6005 & 144889 & 1.37 & K4III         &  1.0 &    2230 &    0.14 & \nodata & \nodata & \nodata & \nodata & \nodata &      22 \\
 6018 & 145328 & 1.01 & K0III--IV     &  1.0 & \nodata & \nodata &      31 &    4.49 &    0.81 &    6925 & 8.98 & \nodata \\
 6046 & 145849 & 1.34 & K3III         &  3.2 &    2150 &     0.6 &       6 & \nodata & \nodata & \nodata & \nodata & \nodata \\
 6148 & 148856 & 0.94 & G7IIIa        &  4.8 & 410.575 &    0.55 & \nodata & \nodata & \nodata & \nodata & \nodata &       2 \\
 6239 & 151627 & 0.87 & G5III         &  4.1 & \nodata & \nodata &       2 &    6.39 &    4.52 &     333 & 12.78 & \nodata \\
 6322 & 153751 & 0.89 & G5III         & 23.0 & 39.4816 &   0.007 & \nodata & \nodata & \nodata & \nodata & \nodata &       7 \\
 6363 & 154732 & 1.09 & K1III         &  1.7 &   790.6 &   0.217 & \nodata & \nodata & \nodata & \nodata & \nodata &      26 \\
 6388 & 155410 & 1.28 & K3III         &  1.0 &  876.25 &   0.609 & \nodata & \nodata & \nodata & \nodata & \nodata &      10 \\
 6577 & 160365 & 0.56 & F6III         & 30.0 & \nodata & \nodata &       3 &   17.92 &   10.35 &    1829 & 35.84 & \nodata \\
      & 160538 & 1.05 & K2III         &  7.2 &   903.8 &   0.072 & \nodata & \nodata & \nodata & \nodata & \nodata &       8 \\
 6790 & 166207 & 1.04 & K0III         &  1.6 & \nodata & \nodata &       2 &    1.66 &    1.17 &     276 & 3.32 & \nodata \\
 6791 & 166208 & 0.91 & G8IIICN-0.3CH &  3.2 &    2017 & 0.378   & \nodata & \nodata & \nodata & \nodata & \nodata &      28 \\
 6853 & 168322 & 0.99 & G9III         &  1.8 & \nodata & \nodata &      22 &    2.76 &    0.59 &    5542 & 5.52 & \nodata \\
 6860 & 168532 & 1.53 & K3III:Ba0.    &  3.9 &  485.45 & 0.359   & \nodata & \nodata & \nodata & \nodata & \nodata &      46 \\
 6886 & 169221 & 1.07 & K1III         &  2.0 & \nodata & \nodata &       2 &    2.82 &    1.99 &   339   & 5.64 & \nodata \\
 7010 & 172424 & 0.96 & G8III         &  1.5 & \nodata & \nodata &       3 &    1.03 &    0.59 &    372  & 2.06 & \nodata \\
 7024 & 172831 & 1.00 & K1III         &  1.3 &   485.3 & 0.209   & \nodata & \nodata & \nodata & \nodata & \nodata &      14 \\
 7125 & 175306 & 1.19 & G9IIIbCN-0.5  & 14.7 & 138.420 & 0.11    & \nodata & \nodata & \nodata & \nodata & \nodata &       2 \\
 7135 & 175515 & 1.04 & K0III         &  1.2 &    2994 & 0.24    & \nodata & \nodata & \nodata & \nodata & \nodata &      12 \\
 7137 & 175535 & 0.90 & G8III         &  2.3 & \nodata & \nodata &       6 &    3.69 &    1.51 &    2579 & 7.38 & \nodata \\
 7176 & 176411 & 1.08 & K1IIICN0.5    &  2.1 &  1270.6 & 0.27    & \nodata & \nodata & \nodata & \nodata & \nodata &      15 \\
 7180 & 176524 & 1.15 & K0III         &  2.1 &  258.48 & 0.21    & \nodata & \nodata & \nodata & \nodata & \nodata &      16 \\
 7208 & 176981 & 1.67 & K2III         &  1.8 & \nodata & \nodata &       2 &    3.91 &    2.77 &     295 &  7.82 & \nodata \\
 7252 & 178208 & 1.27 & K3III         &  1.4 & \nodata & \nodata &       2 &   10.31 &    7.29 &     338 &  20.62 & \nodata \\
 7333 & 181391 & 0.92 & G8III--IV     &  2.8 & 266.544 & 0.833   & \nodata & \nodata & \nodata & \nodata & \nodata &       9 \\
 7413 & 183611 & 1.39 & K5III         &  2.1 & \nodata & \nodata &       2 &    1.71 &    1.21 &     331 & 3.42 & \nodata \\
 7636 & 189322 & 1.13 & G8III         &  1.0 & \nodata & \nodata &       2 &    3.30 &    2.34 &     312 & 6.60 & \nodata \\
 7798 & 194152 & 1.08 & K0IIIv        &  2.1 & 1124.06 & 0.759   & \nodata & \nodata & \nodata & \nodata & \nodata &      31 \\
 7884 & 196574 & 0.95 & G8III         &  3.7 &   205.2 & 0.138   & \nodata & \nodata & \nodata & \nodata & \nodata &      40 \\
 7897 & 196758 & 1.06 & K1III         &  1.8 & \nodata & \nodata &       2 &    2.74 &    1.94 &     377 & 5.48 & \nodata \\
 7901 & 196787 & 1.02 & G9III         &  2.9 & \nodata & \nodata &      12 &    3.02 &    0.87 &    3637 & 6.04 & \nodata \\
 7939 & 197752 & 1.18 & K2III         &  1.5 &    2506 & 0.383   & \nodata & \nodata & \nodata & \nodata & \nodata &      17 \\
 7995 & 198809 & 0.83 & G7IIICN--1    &  4.7 & \nodata & \nodata &       2 &    3.86 &    2.73 &     358 &  7.72 & \nodata \\
      & 199547 & 1.13 & K0III         &  1.0 &    2871 & 0.632   & \nodata & \nodata & \nodata & \nodata & \nodata &      19 \\
 8035 & 199870 & 0.97 & K0IIIbCN--0.5 &  1.2 &   635.1 & 0.44    & \nodata & \nodata & \nodata & \nodata & \nodata &      43 \\
 8078 & 200817 & 0.99 & K0III         &  2.2 & \nodata & \nodata &       2 &    3.98 &    2.81 &     366 &  7.96   & \nodata \\
 8149 & 202951 & 1.65 & K5III         &  4.4 & \nodata & \nodata &       2 &    5.12 &    3.62 &     370 &  10.24   & \nodata \\
 8359 & 208110 & 0.80 & G0IIIs        &  3.3 & \nodata & \nodata &       6 &    3.81 &    1.56 &    1844 &  7.62  & \nodata \\
      & 209813 & 1.08 & K0III         & 21.1 & 24.4284 & 0.01    & \nodata & \nodata & \nodata & \nodata & \nodata &       2 \\
 8442 & 210220 & 0.88 & G6III         &  1.8 & \nodata & \nodata &       2 &    7.03 &    4.97 &     381 &  14.06  & \nodata \\
 8445 & 210289 & 1.62 & K5III         &  2.0 & \nodata & \nodata &       2 &    1.67 &    1.18 &     374 &  3.34   & \nodata \\
 8575 & 213389 & 1.15 & K2III         & 34.4 &  17.755 & 0.02    & \nodata & \nodata & \nodata & \nodata & \nodata &       2 \\
      & 217188 & 1.10 & K0III         &  3.0 & 47.1162 & 0.463   & \nodata & \nodata & \nodata & \nodata & \nodata &       7 \\
 8748 & 217382 & 1.43 & K4III         &  1.0 & \nodata & \nodata &       2 &    1.04 &    0.73 &     334 &  2.08   & \nodata \\
 8819 & 218658 & 0.80 & G2III         &  5.5 &  556.72 & 0.297   & \nodata & \nodata & \nodata & \nodata & \nodata &      46 \\
 8923 & 221115 & 0.94 & G7III         &  1.5 & \nodata & \nodata &       2 &    2.38 &    1.68 &     241 &  4.76   & \nodata \\
 8961 & 222107 & 1.01 & G8III--IV     &  6.9 & 20.5212 & 0.04    & \nodata & \nodata & \nodata & \nodata & \nodata &       2 \\
 8990 & 222682 & 1.24 & K2III         &  1.0 & \nodata & \nodata &       2 &    1.18 &    0.83 &     287 &  2.36   & \nodata \\

\enddata

\tablerefs{1-- Basset 1978; 2-- Batten et al. 1989; 3-- Beavers
and Griffin 1979; 4-- Bertiau 1957; 5-- Carquillat et al. 1983;
6-- Christie 1936; 7-- De Medeiros and Udry 1999; 8-- Fekel et al.
1993; 9-- Franklin 1952; 10-- Griffin 1978; 11-- Griffin 1981a;
12-- Griffin 1981b; 13-- Griffin 1982a; 14-- Griffin 1982b; 15--
Griffin 1982c; 16-- Griffin et al. 1983; 17-- Griffin 1983; 18--
Griffin 1984a; 19-- Griffin 1984b; 20-- Griffin 1985; 21-- Griffin
1986; 22-- Griffin 1988; 23-- Griffin 1990; 24-- Griffin 1991a;
25-- Griffin 1991b; 26-- Griffin 1991c; 27-- Griffin 1991d; 28--
Griffin 1992; 29-- Griffin 1998; 30-- Griffin and Eitter 1999;
31-- Griffin and Eitter 2000; 32-- Griffin et al. 1985; 33--
Griffin et al. 1990; 34-- Griffin et al. 1992; 35-- Griffin and
Gunn 1977; 36-- Griffin and Herbig 1981; 37-- Hall et al. 1995
38-- Jackson et al. 1957; 39-- Lucke and Mayor 1982; 40-- Lucy and
Sweeney 1971; 41-- Luyten 1936; 42-- Morbey and Brosterhus 1974;
43-- Radford and Griffin 1975; 44-- Scardia et al. 2000; 45--
Scarfe 1971; 46-- Scarfe et al. 1983; 47-- Scarfe and Alers 1975;
48-- Strassmeier et al. 1989.}

\end{deluxetable}

\end{document}